\documentclass[reprint,
frontmatterverbose, 
 showpacs,showkeys,
 amsmath,amssymb,
 aps,
]{revtex4-1}

\usepackage{graphicx}
\usepackage{amssymb}
\usepackage{epsfig}
\usepackage{amsmath}
\usepackage{dcolumn}% Align table columns on decimal point
\usepackage{bm}% bold math

\newcommand{\be}{\begin{equation}}
\newcommand{\ee}[1]{\label{#1} \end{equation}}
\newcommand{\ba}{\begin{eqnarray}}
\newcommand{\ea}[1]{\label{#1} \end{eqnarray}}
\newcommand{\nl}{\nonumber \\}

\begin{document}
%\preprint{}
\title{Jet Mass Dependence of Fragmentation\\in Positron-Proton Collisions}
\thanks{These research results were sponsored by the China/Shandong University International Postdoctoral Exchange Program. This paper was also supported by the Hungarian OTKA Grant K104260.}

\author{K. Urmossy}
\email[]{karoly.uermoessy@cern.ch}
\affiliation{Shandong University, 27 Shanda Nanlu, Licheng, Jinan, P.R. China, 250100}
\affiliation{On leave from Wigner RCP, 29--33 Konkoly-Thege Miklos Str., Budapest Hungary, H-1121}
\date{\today}
\begin{abstract}
We propose the characterization of fragmentation functions by the energy fraction $\tilde x$, a hadron takes away from the energy of the jet measured \textit{in the frame co-moving with the jet}. Besides, we propose the usage of the \textit{jet mass as the fragmentation scale} $\tilde Q$. We show that these two Lorentz-invariant variables emerge naturally in a microcanonical ensemble with conserved fourmomentum. Then, we construct a statistical hadronisation model, in which, two features of the hadronic final states in various high-energy reactions (power law spectra and negative-binomial multiplicity distributions) can be connected simply. Finally, we analyse the scale dependence of the parameters of the model (power of the spectrum and mean energy per hadron) in the $\phi^3$ theory. Fitting fragmentation functions in diffractive positron-proton collisions, we obtain a prediction for the jet mass dependence of the hadron multiplicity distribution inside jets.
\end{abstract}
%\keywords{}



\maketitle

\section{Introduction}
\label{sec:intro}

Factorization theorem (FT) \cite{bib:css} in quantum chromo-dynamics (QCD) allows for the approximation of hadron distributions in high-energy processes as a convolution of the cross section of the creation of energetic real partons and of distributions containing details of the soft processes, such as parton distributions (PDFs) in initial-state hadrons and the probabilities of partons to fragment into hadrons (fragmentation functions - FFs). Due to the absence of hadrons in the initial state, electron-positron ($e^+e^-$) annihilations are the best reaction type to study FFs. In the factorised picture, in an $e^+e^-\rightarrow$ 2 jets event at collision energy $\sqrt s$, an on-shell quark (q) and an anti-quark ($\bar q$) are created with momenta (neglecting quark masses) $p_{q/\bar{q}}^\mu = \left(\sqrt s/2,0,0,\pm\sqrt s/2\right)$. Denoting the total momentum by $q^\mu = p_q^\mu + p_{\bar q}^\mu$, the probability $D\left(x,Q^2\right)$ (fragmentation function) of finding a hadron with momentum $p_h$ in the final state can be described by two Lorentz-invariant variables: the energy fraction $x = 2p_h^\mu q_\mu/q^2 = 2p^0_h/\sqrt{s}$ the hadron takes away from that of the (anti-)quark; and the fragmentation scale $Q^2 = q^2 = s$. Measuring the $d\sigma/dx$ hadron distributions in 2-jet events would, thus, provide direct access to the quark-to-hadron fragmentation functions.

A problem with this framework is that an on-shell ``leading'' quark enters the fragmentation process which results in a jet with momentum $P_{jet}^\mu$ and mass $M_{jet}^2 = P_{jet}^2$ which is considerably higher than that of the leading quark, $M_{jet}^2 \gg p_q^2$. As an example, in $e^+e^-$ annihilations at center-of-mass energies $\sqrt s$ = 14--44 GeV \cite{bib:TassoMjet}, the heavy jet mass can be 15--50\% of $\sqrt s/c^2$. In proton-proton (pp) collisions at $\sqrt s$ = 7 TeV \cite{bib:atlasM}, the mass of a jet of transverse momentum $P^{jet}_T$ = 200--600 GeV/c, is typically of order $M_{jet}\sim$ 100 GeV$/c^2$. This is a clear violation of energy-momentum conservation, as the momentum of the final state jet does not coincide with that of the quark which initiated the jet: $P^\mu_{jet} \not= p^\mu_q$. This is a side-effect of a framework, in which, the virtualities of the partons created in the hard process are neglected. However, if the virtualities of the leading (anti-)quark are not neglected in an $e^+e^-\rightarrow$ 2 jets event, energy-momentum conservation only requires that $p_q^\mu = \left(p_q^0,\mathbf{p}_q\right)$ and $p_{\bar{q}}^\mu = \left(\sqrt{s} - p_q^0,-\mathbf{p}_q\right)$. This way, the energies and masses of the heavy and the light jets are different and fluctuate \cite{bib:TassoMjet}.

At this point, three problems emerge:
\begin{itemize}
 \item[$i,$] as the energy of a jet is not $\sqrt s/2$, $x = p_h/\left(\sqrt s/2\right)$ is no-longer the energy fraction a hadron takes away from the jet.
 \item[$ii,$] The energy scale $\tilde Q^2$ at which the fragmentation takes place is not $s$.
 \item[$iii,$] Measured $d\sigma/dx$ distributions in $e^+e^-$ annihilations (e.g. \cite{bib:opal200} and Refs. therein) are mixtures of hadron yields from jets of fluctuating energies, masses and 3-momenta.
\end{itemize}

A reasonable solution to the first issue is to use $\tilde x = 2p^q_\mu p_h^\mu/p_q^2 = 2P^{jet}_\mu p_h^\mu/P_{jet}^2 = 2p_h^{0\,cm}/M_{jet}$, which is in fact the energy fraction the hadron takes away from that of the leading (anti-)quark in the frame co-moving with the jet ($p_h^{0\,cm}$ is the energy of the hadron in the co-moving frame).

As for the fragmentation scale, in theoretical calculations \cite{bib:MLLA1,bib:MLLA2,bib:MLLA3,bib:MLLA4,bib:MLLA5,bib:MLLA6,bib:dEnterria1,bib:dEnterria2} of the distributions of partons inside jets and their scale evolution, the jet opening angle $\theta_c$ and the energy of the leading parton $p_q^0$ are used to define the scale $Q_{th} = \theta_c p_q^0$. There, $Q_{th}$ serves as the upper limit for the transverse momenta of the radiated partons (the transverse width in phasespace) in a small-angle parton evolution process. As we will show in Sec.~\ref{sec:relM}, energy-momentum conservation for the $p_i^\mu = \left(p_i^0,\mathbf{p}_i^T,p_i^z\right)$ momenta of the radiated partons provides a Lorentz-invariant upper limit (unlike $\theta_c p_q^0$) in the transverse direction: $\left|\mathbf{p}_i^T\right| \leq \sqrt{p^2_q}/2$. This limit is valid for large angle radiations as well. Thus, the leading parton's virtuality which coincides with the mass of the final state jet is a natural choice for the fragmentation scale $\tilde Q^2 \sim p_q^2 = M_{jet}^2$. 
 
\textit{For the above reasons, we propose the measurement of the fragmentation function $D\left(\tilde x,\tilde Q^2\right)$ which depends on the newly introduced variables $\tilde x = 2p_h^{0\,cm}/M_{jet}$ and $\tilde Q^2 = M_{jet}^2$. This requires the identification of jets, grouping them into mass bins, and presenting hadron distributions inside each jet in the frame co-moving with the jet.}

Since such datasets are not yet available, we have picked datasets in which, at least, the jets are identified. We have analysed longitudinal and transverse momentum fraction distributions of hadrons inside jets stemming from pp collisions at $\sqrt s$ = 7 TeV \cite{bib:atlasFFpp7TeV} in another paper \cite{bib:UKpp3D}. In that dataset, jets are separated into bins according to their transverse momentum $P^T_{jet}$, however, the jet mass in each $P^T_{jet}$ bin has broad fluctuations \cite{bib:atlasM}. In this paper, we focus on large rapidity gap \textit{$e^+p$} collisions events \cite{bib:H1}, where, hadron yields stemming from the fragmentation of the proton can be separated from yields stemming from current fragmentaion. Though in this dataset, dijet final states are examined, the difference between the energies of the two jets are required to be less than 20\% in the center-of-mass frame of the dijet, so the two jets are kinematically (approximatelly) identical, having momenta $P_{1,2}^\mu = \left(E_{jet},\pm\mathbf{P}_{jet} \right)$. Similarly to the pp case, jets are binned with respect to their energies, while their masses are allowed to fluctuate, though, there is no published data on the distribution of these masses.

In Sec.~\ref{sec:relM}, we show that the variables $\tilde x$ and $\tilde Q$ emerge naturally in a relativistic microcanonical ensemble with conserved fourmomentum. Furtheremore, inclusion of negative-binomial particle multiplicity fluctuations results in \textit{cut-power law} shaped single particle distributions. Both of them have been observed in various types of high-energy ($e^+e^-$, pp and heavy-ion (AA)) collisions \cite{bib:UKpp3D,bib:UKeeFF,bib:UKppFF,bib:UKeeFFstrange,bib:Becattini10,bib:Liu1,bib:Wibig3,bib:FHLiu,bib:Begun3,bib:Begun2,bib:Begun,bib:Beck1,bib:Wilk5,bib:BiroJako,bib:Cley,bib:Khan,bib:Pars,bib:Wong2,bib:Tang2,bib:UKinAuAu,bib:De,bib:BiroResFlukt,bib:UK_WPCF2014,bib:UK_HpT2014,bib:Biro_eeFF}.
In Sec.~\ref{sec:exp}, we fit the cut-power law distribution obtained in Sec.~\ref{sec:relM} to the above described $e^+p$ dataset \cite{bib:H1} on FFs. As we beleive that the correct fragmentation scale is the jet mass, however, the jet mass distributions corresponding to the $E_{jet}$ bins are yet unknown, we fit a characteristic/average mass in case of each dataset in the $E_{jet}$ bins. This way, we obtain the dependence of the parameters of the model on the fragmentation scale $M_{jet}$.

In Sec.~\ref{sec:Q}, we descuss the scale evolution of the parameters using an approximate solution of the DGLAP (Dokshitzer-Gribov-Lipatov-Alterelli-Parisi) equation in the simplest asymptotically free field theory, the $\phi^3$ theory with LO splitting function and 1-loop coupling \cite{bib:UKphi3}.

\section{Statistical Jet Fragmentation}
\label{sec:relM}
Microcanonical statistics has been widely used in the literature in the description of hadronisation \cite{bib:UKpp3D,bib:UKeeFF,bib:UKppFF,bib:UKeeFFstrange,bib:Becattini10,bib:Liu1,bib:Wibig3,bib:FHLiu,bib:Begun3,bib:Begun2,bib:Begun}. Using the microcanonical instead of the canonical ensemble is important, as the energy of a single hadron in a jet can easily be of the order of the total energy of the jet. The advantage of the model presented here is that (after neglecting hadron masses, and using a relativistic ensemble,) we are able to derive simple analytic expressions, while the usage of a non-relativistic ensemble leads to rather complicated results \cite{bib:Begun3}. The disadvantage of neglecting hadron masses is that ratios of total particle multiplicities cannot be 
reproduced as has been done in more complicated simulations \cite{bib:Becattini10,bib:Liu1,bib:Wibig3,bib:FHLiu} taking masses into account.

\subsection{Hadron Distribution in a Single Jet}
\label{sec:1jet}

Generalizing results in \cite{bib:UKppFF,bib:UKeeFF}, we model the created hadrons in a single jet of momentum $P^{\mu}=(E,\mathbf{P})$, mass $M = \sqrt{P_\mu P^\mu}$ and hadron multiplicity $n$ by a microcanonical ensemble. Neglecting hadron masses, writting their momenta as $p_i^\mu = (p_i,\mathbf{p}_i)$, the phasespace corresponding to the jet is 
\be
\Omega_n(P^\mu) \; = \;  \prod_{i=1}^n \int \frac{d^3\mathbf{p}_i}{p^0_i} \, \delta^4\left( \sum_{j=1}^n p^\mu_j - P^\mu \right)\; \propto M^{2n-4} \;.
\ee{eq1}
The evaluation of the integrals in Eq.~(\ref{eq1}) is straightforward in a frame co-moving with the ensemble, and using Fourier transforms: $\delta(x) \propto \int\limits_{-\infty}^\infty d\alpha\, \exp(i\alpha x)$ and $\Omega_n(P^\mu) \; \propto \;  \int d^4\alpha \, \exp\left(i \alpha_\mu P^\mu\right) \varphi^n(\alpha)$ with $\varphi(\alpha) \; = \;  \int \left(d^3\mathbf{p}/p^0\right) \, \exp\left(-i \alpha_\mu p^\mu\right)$.

Thisway, the single-particle distribution is
\ba
f_n(p_\mu) &\;=\;& \frac{\Omega_{n-1}(P^\mu - p^\mu)}{\Omega_n(P^\mu)} \nl
&\;=\;& \frac{(n-1)(n-2)}{\pi M^2} \left(1 - \tilde x \right)^{n-3}  \;,
\ea{eq2}
with $\tilde x = 2 P_\mu\, p^\mu / M^2$ and normalisation condition
\be
1  \;=\; \int \frac{d^3\mathbf{p}}{p^0}f_n(p_\mu)\;.
\ee{eq3}
As $\tilde x\leq 1$ in Eq.~(\ref{eq2}), particle momenta are within an ellipsoid with centre $\mathbf{P}/2$, longer axis $2a = E$ and smaller axis $2b = M$. This way, $\tilde Q = M/2$ is the upper limit for the transverse momenta of hadrons in the jet, similarly to $Q_{th} = \theta_c p_q^0$ which is the upper bound in the transverse direction in case of small-angle parton radiations. In the limit of $|\mathbf{P}|\rightarrow E$, the ellipsoid shrinks, and Eq.~(\ref{eq2}) becomes a one-dimensional distribution of $\tilde x = p^0/E$. As argued in Sec.~\ref{sec:intro}, the variables $\tilde x$ and $\tilde Q$ emerge naturally in a microcanonical ensemble with conserved fourmomentum.

In the canonical limit, where, $n\rightarrow\infty$ while $M = dnT$ is fix ($d=2$ is the effective dimension of the phasespace $d^3\mathbf{p}/p^0 \sim p^d$),  
\be
f_n(p_\mu) \;\rightarrow\;  A \exp\left\lbrace -u_\mu p^\mu/T \right\rbrace  \;,
\ee{eq3b}
which distribution is extensively used in AA collisions for the description of hadrons stemming from the quark-gluon plasma with local flow velocity $u_\mu = P_\mu/M$. \cite{bib:Cley,bib:Tang2,bib:UKinAuAu,bib:De,bib:BiroResFlukt,bib:UK_WPCF2014,bib:UK_HpT2014}.

Interestingly, the statistical distribution, Eq.~(\ref{eq2}) belongs to the family of $x^\alpha (1-x)^\beta$ type parametrisations of FFs most often used in the literature \cite{bib:AKK,bib:Arleo}.

\subsection{Inclusion of Multiplicity Fluctuations}
\label{sec:mult}
Cut-power law momentum distributions along with negative binomial (NBD) multiplicity distributions are common features of hadronic final states in $e^+e^-$, pp and AA collisions as well as in identified jets \cite{bib:UKeeFF,bib:atlasFFpp7TeV_low,bib:UKppFF,bib:PHEn,bib:BiroResFlukt}. Therefore, we average the single particle distribution Eq.~(\ref{eq2}) over the multiplicity fluctuations
\be
\mathcal{P}(n) \;=\; \left(\genfrac{}{}{0pt}{}{n+r-1}{r-1}\right) \tilde{p}^n (1-\tilde{p})^r \;,
\ee{eq4}
and obtain
\ba
p^0\frac{dN}{d^3\mathbf{p}} \;&=&\; \sum \mathcal{P}(n) \, n\, f_n(p_\mu) \,= \nl
&=&  A \left[1 + \frac{q-1}{\tau}\,\tilde x \right]^{-1/(q-1)} - B \;.
\ea{eq5}
To be consistent with notations of other papers in the literature, we introduced the parameters $A = r(r+1)(r+2) [\tilde{p}/(1-\tilde{p})]^3 \;/\; \pi M^2$, $q=1+1/(r+3)$, $\tau = (1-\tilde{p})/[\tilde{p}(r+3)]$ and $B = A[1 + (q-1)/\tau]^{-1/(q-1)}$. Eq.~(\ref{eq5}) follows from Eqs.~(\ref{eq2}) and (\ref{eq4}) using the identity $\sum\limits_{0}^{\infty} \left( \genfrac{}{}{0pt}{}{n+r-1}{r-1}\right) x^n \;=\; (1-x)^{-r}$. The microcanonical nature is manifest in the feature that when particle momenta reach the surface of the ellipsoid ($x = 1$), Eqs.~(\ref{eq2}) and (\ref{eq5}) become zero.

As discussed in \cite{bib:BiroResFlukt}, when $q\rightarrow1$, the multiplicity distribution Eq.~(\ref{eq4}) tends to the Poissonian distribution: $\mathcal{P}(n)\rightarrow (1/\tau)^n \exp\lbrace-1/\tau \rbrace/n!$, while the spectrum Eq.~(\ref{eq5}) tends to Boltzmann-Gibbs $p^0 dN/d^3\mathbf{p}\rightarrow A\exp\lbrace-\tilde x/\tau \rbrace$. Thus, $q-1$ can be viewed as a measure of deviation from the usual canonical distribution. As the mean energy per particle in the co-moving frame
\be
\left\langle p^{0\,cm} \right\rangle \;=\; \frac{d}{1-(d+1)(q-1)} \left(\tau\frac{M}{2}\right)\;
\ee{eq5_2}
tends to the usual equipartition relation $\left\langle p^{0\,cm} \right\rangle \rightarrow d (\tau M/2)$ when $q\rightarrow1$, $T_{eq} = \tau M/2$ may be interpreted as the temperature of the ensemble. Actually, in \cite{bib:Biro_eeFF,bib:Biro_comp}, where Eq.~(\ref{eq5}) is derived from non-extensive thermodynamics with non-additive energy and entropy composition rules, $T_{eq}$ is shown to be the thermodynamical temperature. 

We note here, that Eq.~(\ref{eq5}) may be obtained from other models based on the fluctuations of volume \cite{bib:Begun3,bib:Begun2,bib:Begun}, or temperature \cite{bib:Beck1,bib:Wilk5}; it is also a stationary solution of the Langevin equation with a special kind of multiplicative noise \cite{bib:BiroJako}.

\section{Comparison to $e^+p$ data}
\label{sec:exp}
In this section, we analyse momentum fraction distributions of charged hadrons in jets created in $e^+p \rightarrow e^+XY$ reactions \cite{bib:H1}, where $Y$ is either the intact or excited remnant of the proton, with large rapidity and low transverse momentum. $X$ is of two jets with rapidities $|y_{jet}|\leq1$, and energy difference between the two jets not greater than 20\%. Furtheremore, there is a large rapidity gap between $X$ and $Y$. The $x_p$ distributions were evaluated in the CMS frame of the dijet, with $E_{jet}$ (used in $x_p=p/E_{jet}$) taken to be $E_{jet} = M_{2jet}/2$ defined by $M_{2jet}$ the dijet mass.

Modelling the two jets by two microcanonical ensembles with identical fourmomenta $P^\mu = (E_{jet}, \mathbf{P}_{jet})$, we describe hadronic $x_p$ distributions by Eq.~(\ref{eq5}):
\ba
&&\frac{dN}{dx_p} \;=\;  x_p\,  A \int\limits_0^{\vartheta_c} d\vartheta \sin\vartheta \nl
&&\left\lbrace \left[1 + \frac{q-1}{\tau}\,\frac{E - |\mathbf{P}|\cos\vartheta}{M^2/2} E x_p \right]^{-1/(q-1)} - B\right\rbrace \;, \nl
\ea{eq12}
(we have omitted the subscript 'jet') where $\vartheta_c$ is the opening angle of the jet cone. In total, we have four parameters $A,q,\tau,M$ which we determine by fitting Eq.~(\ref{eq12}) to the datasets corresponding to each $\lbrace E_{jet},\vartheta_c\rbrace$ pairs.

\begin{figure}%[hp]
\begin{center}
\includegraphics[width=0.45\textwidth, height=0.4\textheight]{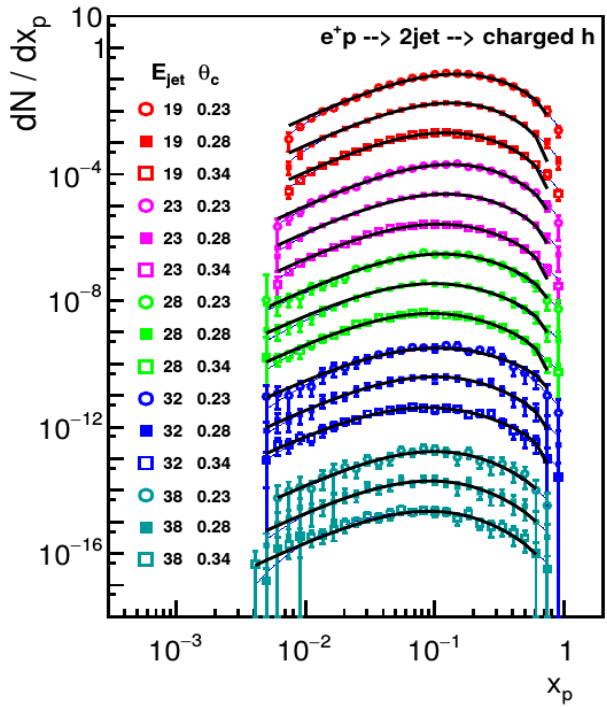} %PLB
\end{center}
\caption{Momentum fraction distributions of charged hadrons stemming from large rapidity gap diffractive $e^+p$ collisions with two jets. Data are published in \cite{bib:H1}. Curves are fits of Eq.~(\ref{eq12}) (thick line) and Eq.~(\ref{eq13}) (thin line). Data are rescaled for visibility.
\label{fig:dNdxp}}
\end{figure}

\begin{figure}%[hp]
\begin{center}
\includegraphics[width=0.45\textwidth, height=0.4\textheight]{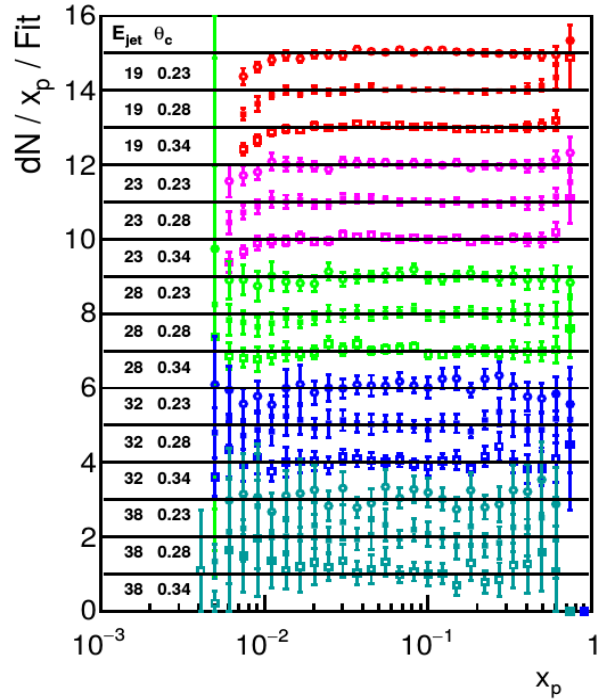} %PLB
\end{center}
\caption{Data over theory plots for Fig.~\ref{fig:dNdxp} using Eq.~(\ref{eq12}). Graphs are shifted by integer numbers for visibility.
\label{fig:dNdxpDperT}}
\end{figure}

As Figs.~\ref{fig:dNdxp}--\ref{fig:dNdxpDperT} shows, Eq.~(\ref{eq12}) reproduces measured data \cite{bib:H1} for $x_p\lessapprox0.7-0.8$. In the case of datasets with $E_{jet}$ = 19 GeV and 23 GeV, there is a discrepancy between our results and both the measured data and the fitted distorted Gaussian (DG) ansatz (proposed in \cite{bib:FWFF}):
\be
\frac{dN}{d\xi}^{DG} =  \frac{\tilde{\mathcal{N}}}{\tilde{\sigma}\sqrt{2\pi}} \exp\left[\frac{\tilde{k}}{8} - \frac{\tilde{s}\tilde{\delta}}{2}- \frac{(2+\tilde{k})\tilde{\delta}^2}{4} + \frac{\tilde{s}\tilde{\delta}^3}{6} + \frac{\tilde{k}\tilde{\delta}^4}{24}\right] ,
\ee{eq13}
where $\xi = -\ln x_p$, $\delta = (\xi - \bar{\xi})/\tilde\sigma$, and $\tilde{\mathcal{N}}, \tilde\sigma, \tilde s, \tilde k$ fit parameters. These discrepancies are around hadron momenta $p\lessapprox$ 150 MeV, which is the low-pT cut-off used in the data analysis. At high $x_p\gtrapprox$ 0.7--0.8, our results underestimate the measured distribution. This might be due to the allowed 20\% energy difference between the two jets in the data analysis. As the energy of the more energetic jet can be 10\% bigger than $M_{2jet}/2$, the $dN/dx_p$ distribution ($x_p = 2p/M_{2jet}$) in that jet might not go to zero at $x_p$ = 1. Thus, measured data might overestimate the real $x_p$ distribution. Another cause might be that we do not take into account the fluctuations of $M_{jet}$, we only fit an average, characteristic value for it. Mass fluctuations may also smear the real value of $x_p$ = 1.


\section{Scale Evolution}
\label{sec:Q}

\begin{figure}%[hp]
\begin{center}
\includegraphics[width=0.45\textwidth, height=0.4\textheight]{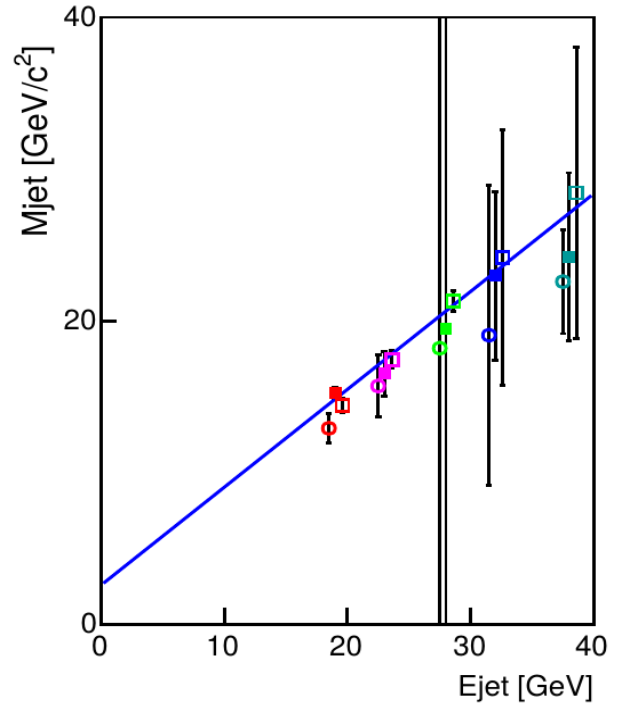} %PLB
\end{center}
\caption{Dependence of the fitted characteristic jet mass on the jet energy and cone opening angle $\theta_c$. Points belonging to a certain $E_{jet}$, but to different $\theta_c$ have been shifted along the horizontal axis for visibility. 
\label{fig:M}}
\end{figure}
The fitted value of the characteristic jet mass shows a growing tendency as a function of $E_{jet}$ (Fig.~\ref{fig:M}). This can be described by a linear function $M_{jet} = M_0 + E_{jet}/E_0$, with $M_0 = (2.6\pm1.4)$ GeV/$c^2$ and $E_0 = (1.6\pm0.2)$ GeV. In the model described in Sec.~\ref{sec:relM}, the mass of the jet defines the width of the jet in phasespace with maximal transverse momentum $p_{Tmax} = M/2$. The values for the maximal transverse momentum obtained from fits of this model are of the order of $p_{Tmax} = E_{jet} \sin\vartheta_c$ which formula (also linear in $E_{jet}$) is used in perturbative analysis \cite{bib:MLLA1,bib:MLLA2,bib:MLLA3,bib:MLLA4,bib:MLLA5,bib:MLLA6,bib:dEnterria1,bib:dEnterria2}. %*** it seems OK, that in our fits M~E as in theor calcs max tra phase spa limit is also ¬E... see plot $E|theta_c$--M

\begin{figure}%[hp]
\begin{center}
\includegraphics[width=0.45\textwidth, height=0.4\textheight]{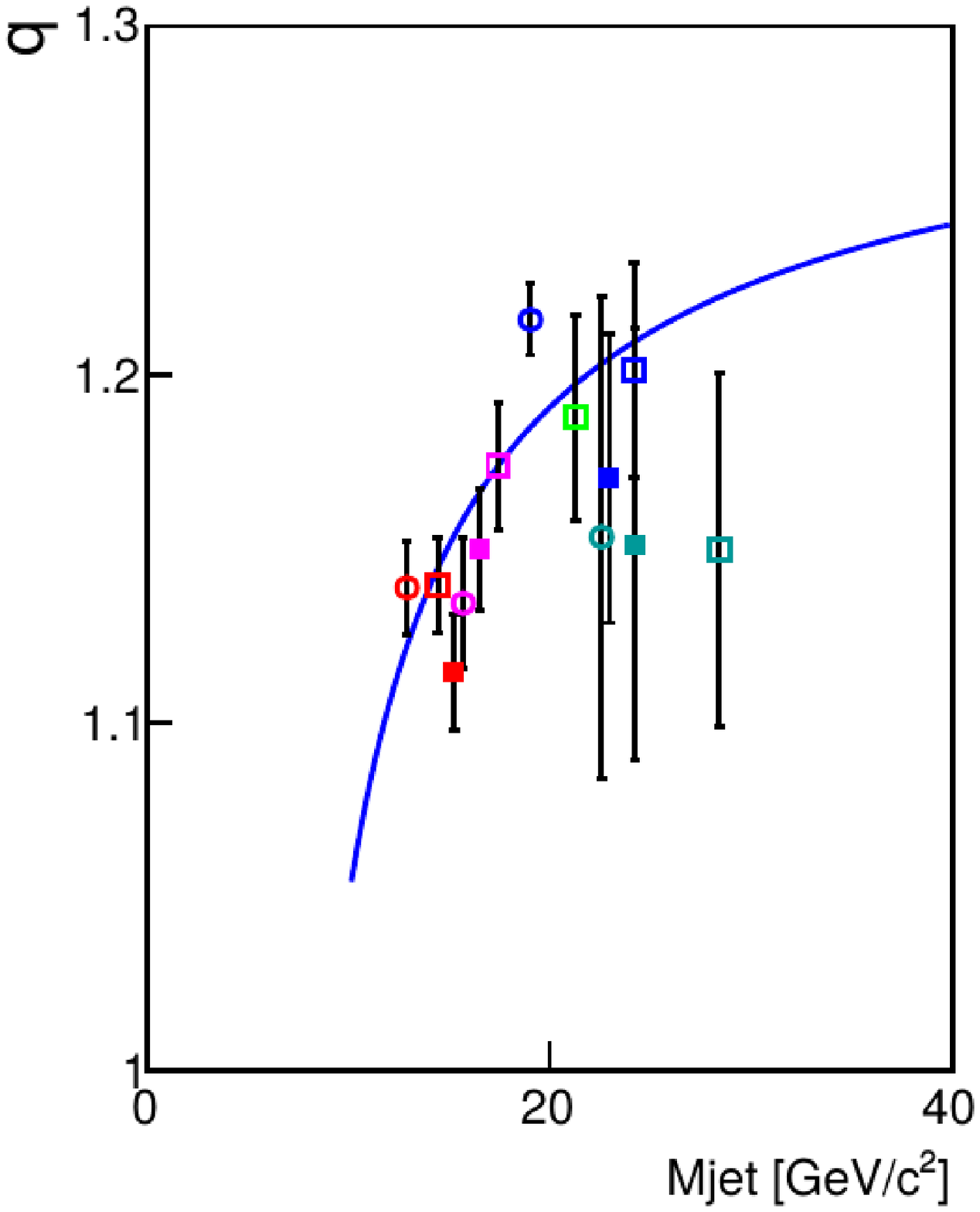} %PLB
\end{center}
\caption{Dependence of the fitted values of the $q$ parameter of Eq.~(\ref{eq12}) on the characteristic jet mass $M_{jet}$. Solid line: $q\left(M^2_{jet}\right)$ from Eq.~(\ref{eq11}) with $Q^2 = M_{jet}^2$.
\label{fig:q}}
\end{figure}

\begin{figure}%[hp]
\begin{center}
\includegraphics[width=0.45\textwidth, height=0.4\textheight]{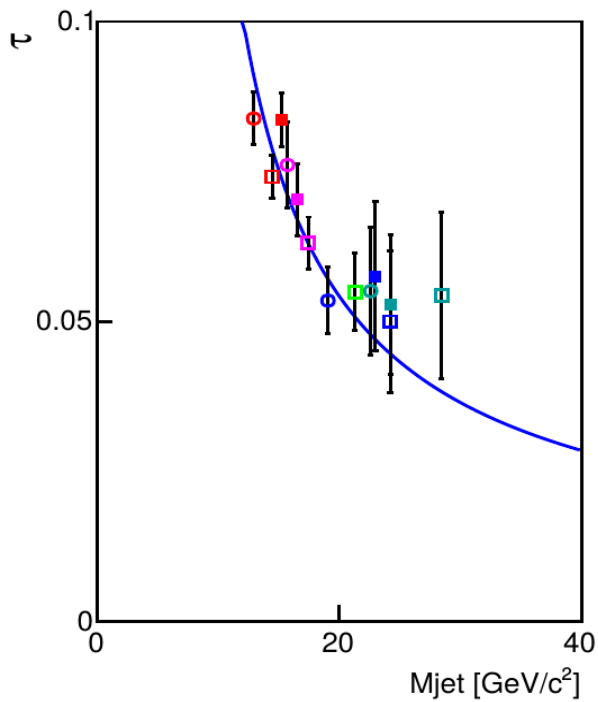} %PLB
\end{center}
\caption{Dependence of the fitted values of the $\tau$ parameter of Eq.~(\ref{eq12}) on the characteristic jet mass $M_{jet}$. Solid line: $\tau\left(M^2_{jet}\right)$ from Eq.~(\ref{eq11}) with $Q^2 = M_{jet}^2$.
\label{fig:T}}
\end{figure}

Dependence of fit parameters $q,\tau$ on our newly proposed fragmentation scale $\tilde Q \propto M_{jet}$ are shown in Figs.~\ref{fig:q}--\ref{fig:T}. As datasets on FFs shown in Fig.~\ref{fig:dNdxp} are binned with respect to the jet energy $E_{jet}$, and in each bin, our newly proposed fragmentation scale, the jet mass most probably fluctuates according to a yet unexamined distribution, it is not possible to make precision tests of QCD (like determination of the strong coupling $\alpha_s\left(M_{jet}^2\right)$). Thus, we derive scale evolution of the $q$ and $\tau$ parameters of the cut-power law shaped FF in (the simplest asymptotically free quantum-field theory) the $\phi^3$ theory. However, the structure of the solution would be very similar in QCD.

Let us define the shape-preserving (approximate) FF
\be
D_a(x,t) \;=\; A(t) \left[1 + \frac{q(t)-1}{\tau(t)}\,x \right]^{-1/[q(t)-1]}\;
\ee{eq9}
which depends on the scale only through its parameters $q(t),\tau(t)$ and $A(t)$, where $t = \ln\left(Q^2/\Lambda^2\right)$, $\Lambda$ is the scale where $g^2(t) = 1/(\beta_0 t)$, the 1-loop coupling of the $\phi^3$ theory diverges and $\beta_0$ is the first coefficient of the beta function of the $\phi^3$ theory. We are intended to obtain the scale $t$ dependence of the parameters using the DGLAP equation in the $\phi^3$ theory, which is
\be
\partial_t D(x,t) \;=\; g^2 \int\limits_x^1 \frac{dz}{z} D\left(\frac{x}{z},t\right) \Pi(z)\;
\ee{eq6}
with the splitting function at LO
\be
\Pi(x) \;=\; x(1-x) - \frac{1}{12}\delta(1-x)\;.
\ee{eq7}
Note that we distinguish between the exact solution $D(x,t)$ and the shape-preserving approximate solution $D_a(x,t)$ for reasons specified later. Eq.~(\ref{eq6}) can be factorized in Mellin space: 
\be
\partial_t \tilde{D}(\omega,t) \;=\; g^2 \tilde{D}\left(\omega,t\right) \tilde{\Pi}(\omega)\;
\ee{eq8_2}
where the Mellin transform of a function $f(x)$ is $\tilde{f}(\omega) = \int\limits_0^1 dx\, x^{\omega-1} f(x)$, so $\tilde{\Pi}(\omega) = 1/[(s+1)(s+2)] - 1/12$. The solution of Eq.~(\ref{eq8_2}) is
\be
\tilde{D}(\omega,t) \;=\; \tilde{D}(\omega,t_0) \exp\lbrace b(t) \tilde{\Pi}(\omega)\rbrace\;
\ee{eq8}
with $b(t) = \int\limits_{t_0}^{t} dt' \, g^2(t') = (1/\beta_0)\ln(t/t_0)$ and $t_0 = \ln\left(Q_0^2/\Lambda^2\right)$ at starting scale $Q_0$. Using the inverse Mellin transform $f(x) \;=\; (1/2\pi)\int\limits_{-\infty}^\infty d\omega\, x^{-i\omega} \tilde{f}(i\omega)$, we obtain the solution in 'real space': 
\be
D(x,t) \;=\; 
\int\limits_x^1 \frac{dz}{z} g\left(z,t\right) D\left(\frac{x}{z},t_0\right)
\ee{phi7}
with the kernel
\ba
g(x,t) &\sim& \delta(x-1) \;+\nl
&+& \sum\limits_{k=1}^\infty \frac{b^k(t)}{k!(k-1)!} \sum\limits_{j=0}^{k-1} \frac{(k-1+j)!}{j!(k-1-j)!} \;\times \nl
&\times& x \ln^{k-1-j}\left[\frac{1}{x}\right] \Big[ (-1)^j + (-1)^k x \Big]
\ea{phi7_2}
and initial function $D(x,t_0)$. 

An inconvenient feature of the solution Eq.~(\ref{phi7}) is that it does not preserve its shape in general. More specifically, substituting  a function $D[\omega,\mathbf{a}(t)]$ which depends on the scale only through its parameters $\mathbf{a}(t)$ into Eq.~(\ref{eq8}), does not in general, lead to a solvable system of equations for $\mathbf{a}(t)$, because the $\omega$ dependence cannot be eliminated in general. For this reason, FFs given in Eqs.~(\ref{eq13}) and (\ref{eq9}) are not exact solutions for the DGLAP equation, though they provide good description of data in a wide range in $x$ (or $\xi$). At this point, we may set up suitable rules to get a sufficient number of equations for the parameters. For example, in \cite{bib:dEnterria1}, the first moments $\left\langle\xi^j\right\rangle$ ($j=0,\dots,5$) of the DG ansatz are required to be equal with the same moments of the exact solution (obtained in QCD at \textit{next-to-modified-leading-log} approximation). In case of Eq.~(\ref{eq9}), 
\ba
&&\tilde{D}_{a}(\omega,t) \approx \int\limits_0^\infty dx\,x^{\omega-1} D_a(x,t) = \nl
&&\;= \frac{A \tau^{\omega-1}}{(q-1)^{\omega-1}} \sum\limits_{j=0}^{\infty} \left(\genfrac{}{}{0pt}{}{\omega-1}{j}\right) \frac{(-1)^{\omega-1-j} \,\tau}{1-(j+1)(q-1)}\nl
\ea{eq7_3}
(we have omitted the $t$ dependence of the parameters). In order to obtain equations for the parameters, we use a simple prescription: let the first three moments of the real and the approximate solutions coincide
\be
\tilde{D}_{a}(\omega,t) \;=\; \tilde{D}(\omega,t)\,, \qquad \omega = \lbrace 1,2,3\rbrace\;.
\ee{eq10}
The $\omega=2$ case is the usual normalisation condition $\int dx\,x D(x,t) = 1$. From Eqs.~(\ref{eq8}), (\ref{eq7_3}) and (\ref{eq10}) we get
\ba
q(t) &=& \frac{(8q_0-12)(t/t_0)^{a_1} - (9q_0-12)(t/t_0)^{-a_2}}{(6q_0-9)(t/t_0)^{a_1} - (6q_0-8)(t/t_0)^{-a_2}}\;, \nl
\tau(t) &=& \frac{\tau_0}{(6q_0-8)(t/t_0)^{-a_2} - (6q_0-9)(t/t_0)^{a_1}} \;,\nl
A(t) &=& [2-q(t)][3-2q(t)]/\tau(t)^2
\ea{eq11}
with $q_0 = q(t_0), \tau_0 = \tau(t_0)$, $a_1 = \tilde{\Pi}(1)/\beta_0$ and $a_2 = \tilde{\Pi}(3)/\beta_0$.


Solid curves in Figs.~\ref{fig:q} and \ref{fig:T} (with $q_0$ = 1.001 and $\tau_0$ = 0.161 at starting scale $Q_0$ = 8.9 GeV/$c^2$ along with $\Lambda$ = 0.1 GeV/$c^2$ and $\beta_0$ = 0.023) show that Eq.~(\ref{eq11}) is consistent with the fitted scale dependence of the parameters of the statistical model with the newly proposed fragmentation scale $\tilde Q = M_{jet}$. As in this paper, we only present a feasibility study on characterizing FFs by the new variables $\tilde x$ and $\tilde Q$, and since the $\phi^3$ theory is not QCD, the actual values of the fitted $\beta_0,\Lambda$ and $Q_0$ are of not much importance. % and fitting the other parameters: $Q_0 = 8.9\pm54, \Lambda = 0.1\pm0.14, \beta_0 = 0.023\pm5.43, q_0 = 1.00101\pm0.07, \tau_0 = 0.161\pm0.004$ 
It is worth to note, however, that in real QCD, in the approximation where the quark/gluon-to-charged hadron fragmentation functions only differ in a normalisation constant, their form 
\be
D_{q/g}^{h}(\omega,t) \;\sim\;D(\omega,t_0)\exp\left\lbrace \int\limits_{t_0}^{t} dt' \gamma(\omega,t') \right\rbrace 
\ee{eq11_2}
is similar to Eq.~(\ref{eq8}), with $\gamma(\omega,t')$ being the anomalous dimension \cite{bib:MLLA1,bib:MLLA2,bib:MLLA3,bib:MLLA4,bib:MLLA5,bib:MLLA6,bib:dEnterria1,bib:dEnterria2}. Thus, the QCD result for the scale dependence of the parameters can be obtained by replacing the $(t/t_0)^{a_{1,2}}$ terms in Eq.~(\ref{eq11}) by the corresponding more complicated functions.

As the parameters of the FF Eq.~(\ref{eq5}) and those of the multiplicity distribution Eq.~(\ref{eq4}) are connected, we have also obtained prediction for the mean multiplicity
\be
\langle n \rangle  \;=\; \frac{4-3q_0}{\tau_0} \left(\frac{t}{t_0}\right)^{-a_2} 
\ee{eq11_3}
and its variance
\be
\left\langle n^2 \right\rangle - \langle n \rangle^2 \;=\; \langle n \rangle \left[ \frac{3-2q_0}{\tau_0} \left(\frac{t}{t_0}\right)^{a_1} - \langle n \rangle  + 1 \right] \;.
\ee{eq11_4}

\section*{Summary}

In this paper, we collect arguments supporting the proposal of describing $D_q^h(\tilde x, \tilde Q^2)$ fragmentation functions (FF) in terms of newly introduced variables: $\tilde x$ being the energy fraction the hadron $h$ takes away from the energy of the jet initiated by parton $q$ in the frame co-moving with the jet; and the fragmentation scale $\tilde Q$ being the mass of the jet. We note that $\tilde Q = M_{jet}$ is Lorentz-invariant, unlike $p_q^0 \theta_c$ (energy of the leading parton $\times$ jet opening angle), usually used in calculations.

In Sec.~\ref{sec:relM}, we show that these variables emerge naturally if we model the creation of hadrons in a jet by a microcanonical ensemble with conserved fourmomentum. The advantage of the model presented in that section, is that it incorporates two experimentally observed phenomena: cut-power law shaped hadron momentum distributions Eq.~(\ref{eq5}) and negative-binomial (NBD) hadron multiplicity distributions ($\mathcal{P}(n)$) Eq.~(\ref{eq4}) inside jets.

Examining the scale dependence of the parameters Eq.~(\ref{eq11}) of the obtained statistical FF in (the simplest asymptotically free quantum field theory) the $\phi^3$ theory, we find that at low scales, the FF is closer to the microcanonical distribution (such configuration corresponds to $q\approx1$), while, for higher scales, the FF evolves to a cut-power function ($q>1$). As our model connects the parameters of the FF and the multiplicity distributions, we also obtain a prediction that at smaller scales $\mathcal{P}(n)$ is closer to a Poissonean distribution with mean multiplicity $\langle n \rangle = 1/\tau$ and for large scales, it becomes a NBD, while the mean multiplicity grows as $\langle n \rangle \sim \ln^a(M_{jet})$ 

We note that the same cut-power law function turned out to describe transverse momentum distributions of hadrons stemming from pp and AA collisions as well, and recently, many empirical formulae have been proposed for the $\sqrt s$ and collision centrality dependence of the $q$ and $\tau$ parameters \cite{bib:De,bib:Wong2,bib:Pars,bib:Khan,bib:BGG_FF}). We believe that understanding the scale dependence of these parameters examining fragmentation will prove useful in the future in the understanding of more complex processes like AA collisions.

\section*{Acknowledgement}
I am grateful for pieces of advice and encouragements I have received from David d'Enterria, Julia Nyiri and Vladimir Anisovich.

\bibliography{UrmossyK_JetMass}

\end{document}